\documentclass[12pt]{article}
\usepackage{axodraw,bbold}

\catcode`@=12
\begin{document}
\thispagestyle{empty}
\begin{flushright} 
UCRHEP-T417\\ 
July 2006\
\end{flushright}
\vspace{0.5in}
\begin{center}
{\LARGE	\bf Supersymmetric Model of Radiative\\ Seesaw Majorana 
Neutrino Masses\\}
\vspace{1.5in}
{\bf Ernest Ma\\}
\vspace{0.2in}
{\sl Physics Department, University of California, Riverside, 
California 92521\\}
\vspace{1.5in}
\end{center}

\begin{abstract}\
The radiative seesaw mechanism proposed recently is minimally extended to 
include supersymmetry in a specific model.  Relevant related issues such 
as leptogenesis and dark matter are discussed.
\end{abstract}

\newpage
\baselineskip 24pt

If the standard model (SM) of particle interactions is extended to include 
three heavy neutral singlet Majorana fermions $N_i~(i=1,2,3)$ and a second 
scalar doublet $\eta = (\eta^+,\eta^0)$ with zero vacuum expectation 
value (VEV), together with an exactly conserved discrete $Z_2$ symmetry 
under which $N_i$ and $\eta$ are odd and all SM particles are even, then 
a radiative seesaw mechanism is obtained \cite{m06-1,kms06,m06-2} for 
small Majorana neutrino masses.  The key is a quartic scalar term
\begin{equation}
\lambda_5 (\Phi^\dagger \eta)^2 + H.c.,
\end{equation}
where $\Phi = (\phi^+,\phi^0)$ is the SM Higgs doublet, which is allowed 
under $Z_2$.  However, such a term is not available in a supersymmetric 
context.  Hence a supersymmetric version of this mechanism is not so 
straightforward.  Nevertheless, it may be accomplished in a minmal 
extension, as shown below.

\begin{table}[htb]
\caption{Particle content of proposed model.}
\begin{center}
\begin{tabular}{|c|c|c|c|}
\hline 
Superfield & $SU(2) \times U(1)$ & $Z_2$ & $Z'_2$ \\ 
\hline
$L_i = (\nu_i,l_i)$ & $(2,-1/2)$ & $-$ & + \\ 
$l^c_i$ & $(1,1)$ & $-$ & + \\ 
$\Phi_1 = (\phi^0_1,\phi^-_1)$ & $(2,-1/2)$ & + & + \\ 
$\Phi_2 = (\phi^+_2,\phi^0_2)$ & $(2,1/2)$ & + & + \\ 
\hline
$N_i$ & $(1,0)$ & $-$ & $-$ \\ 
$\eta_1 = (\eta^0_1,\eta^-_1)$ & $(2,-1/2)$ & + & $-$ \\ 
$\eta_2 = (\eta^+_2,\eta^0_2)$ & $(2,1/2)$ & + & $-$ \\ 
$\chi$ & $(1,0)$ & + & $-$ \\ 
\hline
\end{tabular}
\end{center}
\end{table}

Consider the superfields listed in Table 1.  Whereas $L_i,~l^c_i~(i=1,2,3)$ 
are the usual lepton superfields of the minimal supersymmetric standard 
model (MSSM) and $\Phi_{1,2}$ the usual Higgs superfields, $N_i$ are new 
Majorana lepton superfields, $\eta_{1,2}$ and $\chi$ are new scalar doublet 
and singlet superfields respectively.  The $Z_2 \times Z'_2$ discrete 
symmetry serves to distinguish the three different kinds of doublets 
$L_i$, $\Phi_1$, and $\eta_1$, as well as the two different kinds of 
singlets $N_i$ and $\chi$.  Consequently, the superpotential of this 
model is restricted to be of the form
\begin{eqnarray}
W &=& f_{ij} L_i l^c_j \Phi_1 + h_{ij} L_i N_j \eta_2 + 
\lambda_1 \Phi_1 \eta_2 \chi + \lambda_2 \Phi_2 \eta_1 \chi \nonumber \\ 
&+& \mu_\phi \Phi_1 \Phi_2 + \mu_\eta \eta_1 \eta_2 + {1 \over 2} \mu_\chi 
\chi \chi + {1 \over 2} M_{ij} N_i N_j.
\end{eqnarray}
Since $\eta^0_2$ has zero VEV, $N_j$ are not the Dirac mass partners 
of $\nu_i$ and the canonical seesaw mechanism \cite{seesaw} is not operative. 
However, the Yukawa couplings $h_{ij}$ are available and radiative neutrino 
masses are obtained \cite{m06-1} with the help of $\chi$, as shown in 
Figure 1.

\begin{figure}[htb]
\begin{center}\begin{picture}(500,100)(10,45)
\ArrowLine(70,50)(110,50)
\ArrowLine(150,50)(190,50)
\ArrowLine(150,50)(110,50)
\ArrowLine(230,50)(190,50)
\Text(90,35)[b]{$\nu_i$}
\Text(210,35)[b]{$\nu_j$}
\Text(150,35)[b]{$N_k$}
\Text(150,100)[]{$\chi$}
\Text(105,70)[b]{$\eta^0_2$}
\Text(195,70)[b]{$\eta^0_2$}
\Text(110,116)[b]{$\phi^0_1$}
\Text(190,116)[b]{$\phi^0_1$}
\DashArrowLine(130,85)(115,111){3}
\DashArrowLine(170,85)(185,111){3}
\DashArrowArc(150,50)(40,120,180){3}
\DashArrowArc(150,50)(40,60,100){3}
\DashArrowArcn(150,50)(40,60,0){3}
\DashArrowArcn(150,50)(40,120,80){3}

\ArrowLine(270,50)(310,50)
\DashArrowLine(352,50)(390,50){3}
\DashArrowLine(348,50)(310,50){3}
\ArrowLine(430,50)(390,50)
\Text(290,35)[b]{$\nu_i$}
\Text(410,35)[b]{$\nu_j$}
\Text(350,32)[b]{$\tilde N_k$}
\Text(350,100)[]{$\tilde \chi$}
\Text(305,70)[b]{$\tilde \eta^0_2$}
\Text(395,70)[b]{$\tilde \eta^0_2$}
\Text(310,116)[b]{$\phi^0_1$}
\Text(390,116)[b]{$\phi^0_1$}
\DashArrowLine(330,85)(315,111){3}
\DashArrowLine(370,85)(385,111){3}
\ArrowArc(350,50)(40,120,180)
\ArrowArc(350,50)(40,60,100)
\ArrowArcn(350,50)(40,60,0)
\ArrowArcn(350,50)(40,120,80)

\end{picture}
\end{center}
\caption[]{One-loop radiative contributions to neutrino mass.}
\end{figure}
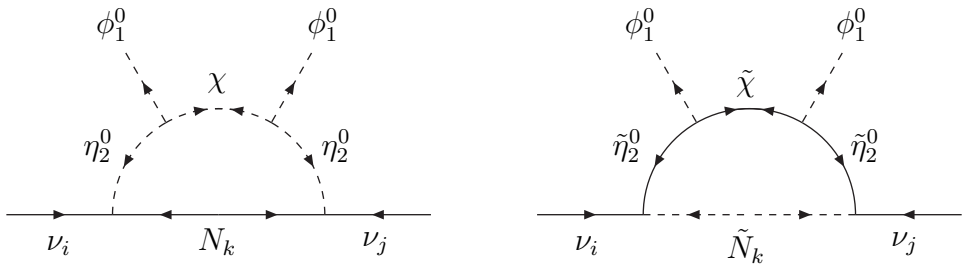

There are two important consequences of this supersymmetric radiative 
seesaw mechanism.  One is leptogenesis \cite{lg}.  The decay of 
the lightest $N_i$ into $L \eta_2$ and their antiparticles will generate 
a lepton asymmetry which gets converted into the observed baryon asymmetry 
of the Universe through sphalerons at the electroweak phase transition 
\cite{krs}.  On the other hand, the formula for neutrino mass is 
suppressed by at least a loop factor of $16 \pi^2$ as compared to that 
of the canonical seesaw.  This means that the Davidson-Ibarra bound \cite{di} 
on leptogenesis is reduced by at least two orders of magnitude and the 
lightest $N_i$ needs to be no heavier than about $10^7$ GeV, allowing it 
to be comfortably below a possible gravitino bound from the reheating of 
the Universe after inflation \cite{kmy06}.

The other is dark matter \cite{dm_rev}.  As in the MSSM, $R-$parity is 
conserved in this model.  The lightest particle with $R=-1$ is stable and 
a candidate for the dark matter of the Universe.  Similarly, the lightest 
particle odd under $Z'_2$ is also stable.  In fact, consider the three 
lightest particles with $(R,Z'_2) = (-,+)$, $(+,-)$, and $(-,-)$ 
respectively. If one is heavier than the other two combined, then the latter 
are the two components of dark matter.  If not, then all three contribute. 
In other words, dark matter may not be as boring as usually assumed. 
It may consist of a rich variety of different stable particles.

Except for $N_i$, the new particles of this model, i.e. $\eta_{1,2}$ 
and $\chi$, are expected to be at the TeV scale and should be observable 
at the forthcoming Large Hadron Collider (LHC).

This work was supported in part by the U.~S.~Department of Energy under Grant 
No.~DE-FG03-94ER40837.

\newpage
\bibliographystyle{unsrt}

\end{document}